\documentclass[12pt]{article}
\pdfoutput=1 
\usepackage{graphicx,amsmath}
\usepackage{units}
\usepackage{color}
\usepackage{float}

\newlength{\dinwidth}
\newlength{\dinmargin}
\def\be{\begin{equation}}   
\def\ee{\end{equation}}  
\def\bea{\begin{eqnarray}}                      
\def\eea{\end{eqnarray}}
\def\ch1{$\chi(1^+)$}
\def\lapproxeq{\lower .7ex\hbox{$\;\stackrel{\textstyle                                                    
<}{\sim}\;$}}                                                    
\def\gapproxeq{\lower .7ex\hbox{$\;\stackrel{\textstyle                                                    
>}{\sim}\;$}}

\begin{document}

\begin{flushright}                                                    
IPPP/16/98  \\                                                    
\today \\                                                    
\end{flushright} 

\vspace*{0.5cm}

\begin{center}

{\Large \bf Scale dependence of open $c\bar{c}$ and  $b\bar{b}$ production} \\
\vspace{0.5cm}
{\Large \bf in the low $x$ region}\\
\vspace{1.0cm}

E.G. de Oliveira$^{a,b}$, A.D. Martin$^b$ and  M.G. Ryskin$^{b,c}$\\ 

\vspace{0.5cm}
$^a$ {\it Departamento de F\'{i}sica, CFM, Universidade Federal de Santa
Catarina, C.P. 476, CEP 88.040-900, Florian\'opolis, SC, Brazil}\\    
$^b$ {\it Institute for Particle Physics Phenomenology, University of Durham, Durham, DH1 3LE } \\
$^c$ {\it Petersburg Nuclear Physics Institute, NRC Kurchatov Institute, Gatchina, St.~Petersburg, 188300, Russia } \\ 

\begin{abstract}
The `optimal' factorization scale $\mu_0$ is calculated for open heavy quark production.
We find that the optimal value is $\mu_F=\mu_0\simeq 0.85\sqrt{p^2_T+m_Q^2} $; a choice which allows us to resum the double-logarithmic, $(\alpha_s\ln\mu^2_F\ln(1/x))^n$ corrections (enhanced at LHC energies by large values of $\ln(1/x)$) and to move them into the incoming parton distributions,
PDF$(x,\mu_0^2)$.
 Besides this result for the single inclusive cross section (corresponding to an observed heavy quark of transverse momentum $p_T$), we also determined the scale for processes where the acoplanarity can be measured; that is, events where the azimuthal angle between the quark and the antiquark may be determined experimentally.  
 Moreover, we discuss the important role played by the $2\to 2$ subprocesses, $gg\to Q\bar{Q}$ at NLO and higher orders.
 In summary, we achieve a better stability of the QCD calculations, so that the data on $c\bar{c}$ and $b\bar{b}$ production can be used to further constrain the gluons in the small $x$, relatively low scale, domain, where the uncertainties of the global analyses are large at present.

\end{abstract}

\end{center}
\vspace{0.5cm}

\section{Introduction}

The present global PDF analyses (e.g.\ NNPDF3.0~\cite{Ball:2014uwa},
MMHT2014~\cite{Harland-Lang:2014zoa}, CT14~\cite{Dulat:2015mca}) find
that there is a large uncertainty in the low $x$ behaviour of the
gluon distribution. There is a lack of appropriate very low $x$ data, particularly
at low scales. 
However, recently measurements on open charm and open beauty in the forward direction have been presented by the LHCb collaboration \cite{LHCbcc7,LHCbcc13,LHCbcc5,LHCbbb}; moreover,
the ATLAS collaboration has measured open charm production in the central rapidity region~\cite{ATLAS}. These data sample the gluon distribution at rather low $x$: namely in the domain $10^{-5} \lapproxeq x \lapproxeq 10^{-4}$. A discussion of the data in terms of existing global PDFs has been presented in \cite{Gauld, Cacciari}, and they have been incorporated in a fit with the HERA deep inelastic data in \cite{Blumlein}.

In the ideal case it would be good to have such data where both the heavy quark and the heavy antiquark were measured, since when we observe only one quark (one heavy hadron) the value of $x$ that is probed is smeared out over an order of magnitude by the unknown momentum of the unobserved quark in the $Q{\bar Q}$-pair \cite{kt,Gauld}, where $Q\equiv c,b$, see e.g. Fig.1 in  ~\cite{Gauld}.  Nevertheless, even measurements of the inclusive cross section of one heavy quark can be used to check and further constrain the existing PDFs.

Another problem, which was emphasized in ~\cite{Cacciari}, is that the QCD prediction  at NLO level strongly depends on the factorization scale, $\mu_F$, assumed in the calculation.  We might expect that the major source of the
strong $\mu_F$ dependence arises because in the DGLAP evolution of low
$x$ PDFs the probability of emitting a new gluon is strongly
enhanced by the large value of $\ln (1/x)$.
Indeed, the mean number of gluons in the interval $\Delta \ln \mu_F^2$
is~\cite{Dokshitzer:1978hw} 
\be
\langle n \rangle \;\simeq\; \frac{\alpha_sN_C}{\pi}\; \ln
(1/x)\;\Delta \ln \mu_F^2\,, 
\label{eq:n}
\ee
leading to a value of $\langle n \rangle$ up to about 8,  for the case
$\ln (1/x)\sim 8$ with the usual $\mu_F$ scale variation interval
from $\mu_F/2$ to $2\mu_F$.  In contrast, the NLO coefficient
function allows for the emission of only {\it one} gluon.
Therefore we cannot expect compensation between the contributions
coming from the PDF and the coefficient function as we vary the scale
$\mu_F$.
It was shown in \cite{DY,Upsilon} that this strong double-logarithmic part of the scale dependence can be successfully resummed by choosing an appropriate scale, $\mu_0$, in the PDF convoluted with the LO hard matrix element, which in our case is ${\cal M}(gg\to Q{\bar Q})$.

The outline of the paper is as follows. In Section~\ref{sec:two} we recall the method of performing the resummation to determine the optimal scale $\mu_0$.  In Section~\ref{sec:renorm} we justify choosing the renormalization scale equal to the factorization scale. Then in Section~\ref{sec:opt41} we use the procedure discussed in Section~\ref{sec:two}, to resum the ln$(1/x)$ terms so as to determine the optimum factorization scale, $\mu_0$.  Unfortunately for heavy $Q\bar{Q}$ production (unlike the Drell-Yan process) a large sensitivity to the choice of scale remains. In Section~\ref{sec:2to2} we identify the source of the problem to be the  important $2\to 2$ (that is $gg\to Q\bar{Q}$) diagrams at NLO and higher orders. We argue that it is possible to also resum these diagrams. We then find the scale sensitivity is reduced. It would be advantageous if both heavy mesons (arising from  $Q$ and $\bar{Q}$) could be measured experimentally, but, at present, the statistics are limited. However, a possibility to circumvent this problem is discussed in Section~\ref{sec:azim}.   In Section~\ref{sec:data} we return to open single inclusive  $c\bar{c}$ and  $b\bar{b}$ production and compare the QCD predictions with the optimal scale with LHC data; and are able to make an observation about the gluon PDF at low $x$. In Section~\ref{sec:conc} we present our conclusion.

\section{Way to choose the optimum factorization scale \label{sec:two}}

Here we recall the procedure proposed in \cite{DY,Upsilon}, which provides a reduction in the sensitivity to the choice of factorization scale by resumming the enhanced double-logarithmic contributions from a knowledge of the NLO contribution.  The cross section for open heavy quark production at LO + NLO at factorization scale $\mu_f$ may be expressed in the form\footnote{For
  ease of understanding we omit the parton labels $a=g,q$ on the
  quantities in~(\ref{eq:sta}) and the following equations. The matrix
  form of the equations is implied.} 
\begin{equation}
\sigma^{(0)}(\mu_f) + \sigma^{(1)}(\mu_f) ~=~ \alpha_s^2\left[{\rm PDF}(\mu_f)\otimes C^{(0)} \otimes {\rm PDF}(\mu_f) +
{\rm PDF}(\mu_f)\otimes\alpha_s C^{(1)}(\mu_f) \otimes {\rm PDF}(\mu_f)\right]\,, 
\label{eq:sta}
\end{equation}
where the coefficient function
$C^{(0)}$ does not depend on the factorisation scale, while the $\mu_f$ dependence of the NLO coefficient function arises since we have to subtract from the NLO diagrams the  part already generated by LO evolution. 

We are
free to evaluate the LO contribution at a different scale $\mu_F$,
since the resulting effect can be {\it compensated} by changes in the
NLO coefficient function, which then also becomes dependent on
$\mu_F$. In this way eq.~(\ref{eq:sta}) becomes 
\begin{equation}
\sigma^{(0)}(\mu_f) + \sigma^{(1)}(\mu_f) ~=~\alpha_s^2\left[{\rm PDF}(\mu_F)\otimes C^{(0)} \otimes {\rm PDF}(\mu_F) +
{\rm PDF}(\mu_f)\otimes\alpha_s C_{\rm rem}^{(1)}(\mu_F) \otimes {\rm PDF}(\mu_f)\right]\,.  
\label{eq:stab1}
\end{equation}
Here the first $\alpha_s$ correction $C^{(1)}_{\rm rem}(\mu_F)\equiv C^{(1)}(\mu_f=\mu_F)$ is calculated now at the scale $\mu_F$ used for the LO term, and not at the scale $\mu_f$ corresponding to the cross section on the left hand side of the formula. Since it is the correction which {\em remains} after the factorization scale in the LO part is fixed, we denote it $C_{\rm rem}^{(1)}(\mu_F)$.
Note that although the first and second terms on the right hand side
depend on $\mu_F$, their sum, however, does not (to ${\cal O}(\alpha_s^4)$), and
is equal to the full LO+NLO cross section calculated at the factorization
scale $\mu_f$.

Originally the NLO coefficient functions $C^{(1)}$ are calculated
from Feynman diagrams which are independent of the factorization
scale. How does the $\mu_F$ dependence of $C^{(1)}_{\rm rem}$
in~(\ref{eq:stab1}) actually arise? It occurs because we must subtract 
from $C^{(1)}$ the $\alpha_s$ term which was already included in the
LO contribution. Since the LO contribution
was calculated up to some scale $\mu_F$ the value of $C^{(1)}$ after
the subtraction depends on the value $\mu_F$ chosen for the LO
component. The change of scale of the LO contribution from $\mu_f$ to
$\mu_F$ also means we have had to change the factorisation scale which
enters the coefficient function $C^{(1)}$ from $\mu_f$ to $\mu_F$. The
effect of this scale change is driven by the LO DGLAP evolution, which
is given by
\be
\sigma^{(0)}(\mu_F)~=~
\alpha_s^2~{\rm PDF}(\mu_f)\otimes \left(C^{(0)} +\frac{\alpha_s}{2\pi}
    \ln\left(\frac{\mu_F^2}{\mu_f^2}\right) (P_{\rm left}\otimes C^{(0)} +C^{(0)} \otimes P_{\rm right})\right)
\otimes {\rm PDF}(\mu_f)\,,
\label{eq:dg}
\ee
where $P_{\rm left}$ and $P_{\rm right}$ denote DGLAP splitting functions acting on the PDFs to the left and right respectively. That is, by choosing
to evaluate $\sigma^{(0)}$ at scale $\mu_F$ we have moved the part of the NLO
(i.e. $\alpha_s$) corrections given by the last term of~(\ref{eq:dg})
from the NLO to the LO part of the cross section. In this way $C^{(1)}$
becomes the remaining $\mu_F$-dependent coefficient function
$C^{(1)}_{\rm rem}(\mu_F)$ of~(\ref{eq:stab1}).
\begin{figure} [h]
\begin{center}
 \includegraphics[clip=true,trim=1.0cm 6.0cm 0.0cm 0.0cm,width=13.0cm]{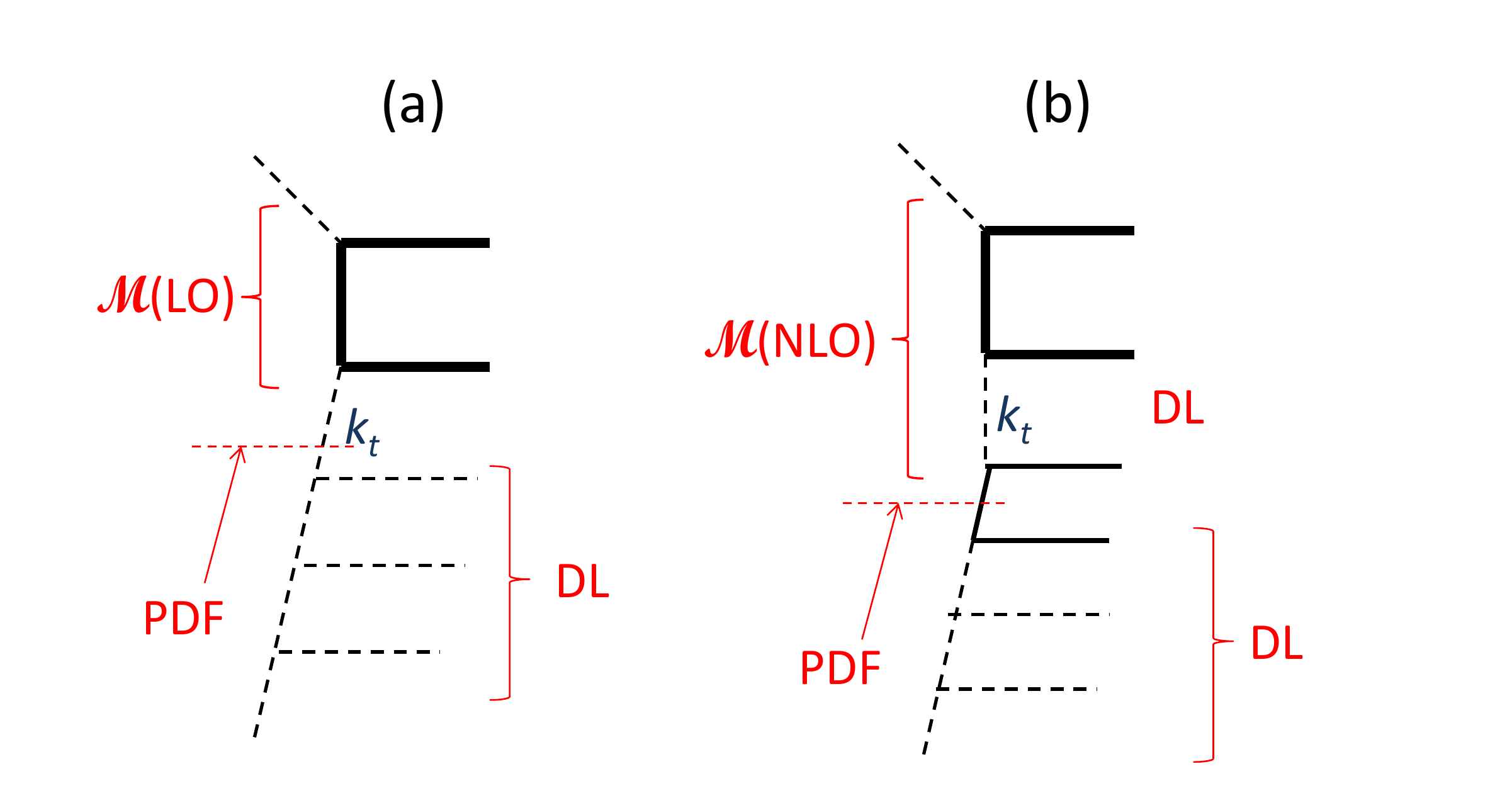} 
%\includegraphics[width=1.0\textwidth]{toLO}
%\vspace{-5.0cm} 
\caption{\sf Heavy quark production at (a) LO via the $gg\to Q{\bar Q}$ subprocess, and (b) via the $gq\to Q{\bar Q}q$ subprocess. The diagrams with $s$-channel gluons $(g^*\to Q{\bar Q})$ are not shown for simplicity. Moreover, only the PDF below the hard matrix element, ${\cal M}$, is shown. Note that the double logarithmic (DL) integral in the NLO matrix element is exactly the same as in the first gluon cell below the LO matrix element. In both cases the ultraviolet convergence is provided by the $k_t$ dependence of the respective matrix element. Therefore it is possible to move the large DL contribution from the coefficient function $C^{(1)}$ of the NLO term, diagram (b), to the PDF term in the LO diagram (a) by choosing an appropriate value of $\mu_F$, and in this way to resum all the higher order DL contributions in the PDFs$(\mu_F)$ of the LO diagram (a).  }
\label{fig:toLO}
\end{center}
\end{figure}
The idea is to choose a scale $\mu_F=\mu_0$ such that the remaining NLO term does not contain the double-logarithmic  $(\alpha_s{\rm ln}( \mu_F){\rm ln}(1/x))^n$ contributions. It is impossible to nullify the whole NLO contribution since the function $C^{(1)}(\mu)$ depends also on other variables; in particular, it depends on the mass, $\hat s$, of the system produced by the hard matrix element. On the other hand we can choose such a value of $\mu$ which makes $C^{(1)}(\mu,\hat s)=0$ in the limit of large $\hat s \gg m^2_Q$. Recall that the $\ln(1/x)$ factor arises in the NLO after the convolution of the large $\hat s$ asymptotics of hard subprocess cross section with the incoming parton low-$x$ distributions satisfying
\be
xq(x)\to {\rm constant}~~~~ {\rm or} ~~~~ xg(x)\to {\rm constant}. 
\ee

At NLO level the change $\mu_f$ to $\mu_F$ is irrelevant: eq.(\ref{eq:stab1}) is an identity (it just changes the higher order terms).  However, in this way we simultaneously resum all the higher order double-logarithmic contributions in the PDFs$(\mu_F)$ of the LO part. As a result we are able to suppress the scale dependence caused by large values of log$(1/x)$.

Thus the choice of $\mu=\mu_0$, which nullifies $C^{(1)}$ at large $\hat s \gg m^2_Q$,
excludes the Double Log (DL), $\alpha_s\ln\mu_F^2\ln(1/x)$, contribution from the NLO correction by  resumming the series of double-logarithmic terms in the PDFs, which are then convoluted with the LO coefficient functions. To find the appropriate value of $\mu_0$ we must choose the NLO subprocess driven by the same ladder-type diagrams (in the axial gauge) as the ladder diagrams that describe LO DGLAP evolution. The appropriate subprocess is gluon-light quark fusion, $gq\to Q{\bar Q}q$. In the high energy limit, where the subprocess energy satisfies ${\hat s}(gq)\gg m^2_Q$, the cross section described by this subprocess contains double-logarithmic terms ~log$(\mu^2_F/\mu^2_0)$ log$({\hat s}/m^2_Q)$. The subprocess $gq\to Q{\bar Q}q$ is contained in the sketch of  Fig.~\ref{fig:toLO}(b), where it is shown pictorially how the enhanced double-logarithmic terms are transferred to the PDFs in the LO term.

\subsection{Extension to higher orders}
We note that, in general, this decomposition can be continued to higher order.  For example, if the NNLO contribution is known, then we will have three scales: $\mu_f,~\mu_F=\mu_0$ and $\mu_1$,
\bea
\sigma^{(0)}(\mu_f) + \sigma^{(1)}(\mu_f) + \sigma^{(2)}(\mu_f)~~=~~\alpha_s^2 ~[~{\rm PDF}(\mu_0)\otimes C^{(0)} \otimes {\rm PDF}(\mu_0) ~+~ ~~~~~~~~~~~~~~~~~~~~~~~\nonumber\\
~~~~~~~~{\rm PDF}(\mu_1)\otimes\alpha_s C_{\rm rem}^{(1)}(\mu_0) \otimes {\rm PDF}(\mu_1) ~+~ {\rm PDF}(\mu_f)\otimes\alpha^2_s C_{\rm rem}^{(2)}(\mu_0,\mu_1) \otimes {\rm PDF}(\mu_f)~]~\,,  
\label{eq:stab11}
\eea
where the scale $\mu_1$ is chosen to nullify the final term in the small $x$ limit.

In fact in Section \ref{sec:2to2}  we will use this equation to include the important $2\to 2$ (that is, $gg\to Q\bar{Q})$ subprocess at NLO and higher orders. We will show reasons why the scale choice $\mu_1=\mu_0$ will give a good approximation for the resummation of these higher-order $2\to 2$ contributions.

\subsection{Comparison with $k_t$ factorization}
\begin{figure} [h]
\begin{center}
\vspace{-0.0cm}
\includegraphics[clip=true,trim=0.0cm 11.0cm 0.0cm 0.0cm,width=13.0cm]{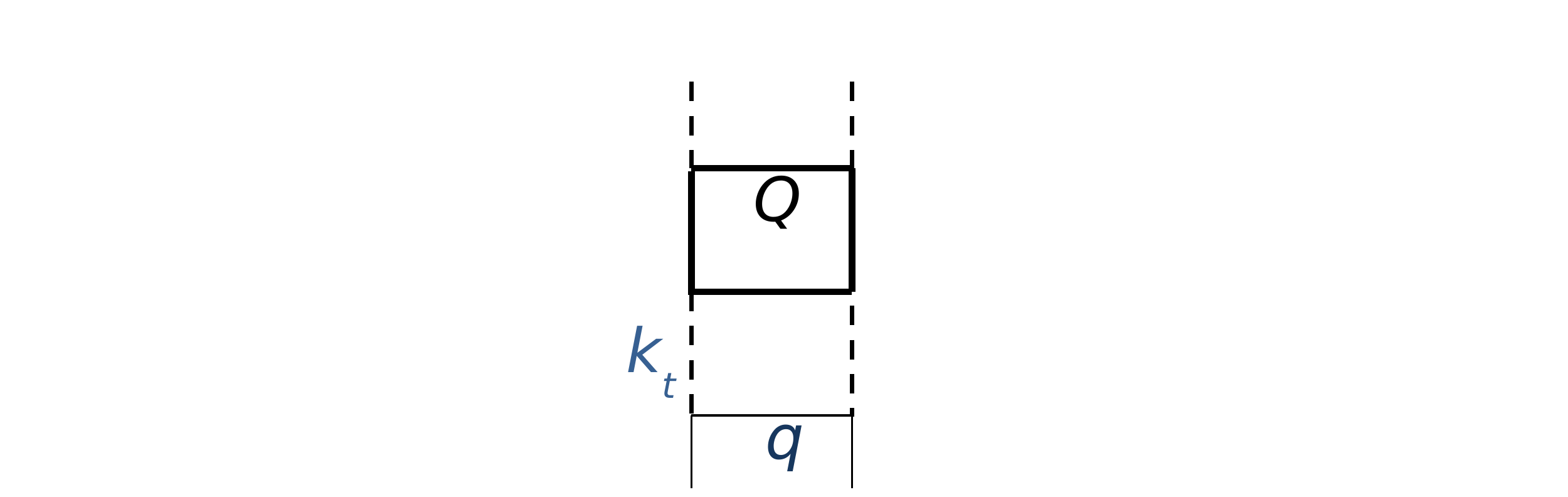}
%\vspace{-7.5cm}
\caption{\sf The diagram for $AA^*$, where $A$ is the amplitude for the subprocess $gg\to Q{\bar Q}q$ shown in Fig.~\ref{fig:toLO}(b). However, in the $k_t$ factorization approach $k_t$ is integrated over and the effective upper limit of the convergent integral essentially plays the role of the appropriate factorization scale. }
\label{fig:kt}
\end{center}
\end{figure}

The approach we have introduced is based on collinear factorization. However, actually it is close in spirit to the
$k_t$-factorization method. Indeed, there, the value of the 
factorization scale is driven by the structure of the integral over $k_t$, see Fig.~\ref{fig:kt}.
In the $k_t$-factorization approach this $k_t$
integral is written explicitly, while the parton distribution {\it
  unintegrated} over $k_t$ is generated by the last step of the DGLAP
evolution, similar to the prescription proposed in
Refs.~\cite{KimbMR,WMR}. Then, using the known NLO result, we account
for the {\em exact} $k_t$ integration in the last cell adjacent to the
LO hard matrix element. This hard matrix element $\cal{M}$, provides the convergence of the
integral at large $k_t$. In this way it puts an effective upper limit
of the $k_t$ integral, which plays the role of an appropriate
factorization scale. 

\section{The renormalization scale $\mu_R$   \label{sec:renorm}}

Besides the factorization scale, the QCD prediction, truncated at NLO, strongly depends on the renormalization scale $\mu_R$, since the LO term is already 
proportional to $\alpha^2_s(\mu_R)$. Let us discuss the possible choice of $\mu_R$. First, it is reasonable to have $\mu_R\gapproxeq \mu_F$, since we expect all the contributions with virtualities less than $\mu_F$ to be included in the PDFs, while those larger than $\mu_F$ to be assigned to the hard matrix element. This is in line with the fact that the current scale of the QCD coupling increases monotonically during the DGLAP evolution. So the coupling responsible for heavy quark production should have a scale $\mu_R$ equal to, or larger than, that in the evolution.

Another argument is based on the BLM prescription \cite{BLM},  which says that all the contributions proportional to $\beta_0=11-\frac{2}{3}n_f$ should be assigned to $\alpha_s$ by choosing an appropriate scale $\mu_R$. 
A good way to trace the $\beta_0$ contribution is to calculate the LO term generated by a new type of light quark, so $n_f\to n_f +1$. Note that the new quark-loop insertion appears twice in the calculation. The part with scales $\mu<\mu_F$ is generated by the virtual ($\propto \delta(1-z)$) component of the LO splitting during DGLAP evolution, while the part with $\mu>\mu_R$ accounts for the running $\alpha_s$ behaviour obtained after the regularization of the ultraviolet divergence. In order  not to miss some contribution and to avoid double counting we take the renormalization scale $\mu_R=\mu_F$. The argument for this choice was made in more detail in~\cite{HKR} for the QED case.

%  At LO the $gg\to Q{\bar Q}$ subprocess is described by diagrams (a) and (b) of Fig.~\ref{fig:blm}. In diagram (a) we can include a new light quark loop only in the incoming gluon propagators; but when these gluons are on-mass-shell this contribution is cancelled by renormalization. For diagram (b) the light quark loop may be included in the $s$-channel gluon propagator or form a loop in the triple gluon vertex. Then the appropriate scale is $\mu_R=m(Q{\bar Q})$.  The same estimate can be obtained from diagram (a) if we consider this diagram in terms of dispersion relations (that is,
% old-fashioned-perturbation-theory(OFPT)), giving diagram (c).
% Indeed, in this case the incoming gluons become heavy and go off-mass-shell. Thus it looks reasonable to start with
%\be
%(\mu_R)_0~=~ 2\sqrt{p_T^2+m_Q^2},
%\ee
%and to study the dependence of the prediction for the cross section to variations of the renormalization scale around this value.

Of course, all these are only the arguments why we expect $\mu_R=\mu_F$, and are not a proof. Formally we can only say that we expect $\mu_R$ to be of the order of $\mu_F$. Thus  there could be further uncertainty in the scale dependence of the predictions due to the possibility that $\mu_R \ne \mu_F$. However, based on these arguments, below we study the factorization scale dependence using the renormalization scale $\mu_R=\mu_0$.

We emphasize (see also \cite{Cacciari}) that the  renormalisation scale dependence affects just the normalization of cross section, but not its energy behaviour. It 
is cancelled in the ratio of the cross sections measured at the LHC energy of 7 (or 8 or 5) TeV to that at 13 TeV or in the ratio of the cross sections obtained at different rapidities. Thus these ratios will probe the low $x$ dependence of the gluons at scale $\mu_F=\mu_0$ essentially without any uncertainties due to possible variations of the $\mu_R$ scale.

\section{Sensitivity of predictions to the factorization scale}

Here we implement the proposals of eqs. (\ref{eq:stab1}) and (\ref{eq:stab11}) in an attempt to reduce the factorization scale dependence of the QCD predictions for $Q\bar{Q}$ production in high-energy $p\bar{p}$ collisions.
Note, however, that calculating the NLO contribution of the diagram in Fig.~\ref{fig:toLO}, we have integrated over the momenta of other particles; in particular, over the transverse momentum, $-k_t$, of the light quark.
% Since we are going to consider the upper (heavy quark) box in Fig.~\ref{fig:toLO} as the `hard' subprocess, and would like to keep the DGLAP $k_t$ ordering, we put an additional cut  $-|k_t|<|p_T|$; otherwise the lower part of diagram, $qQ$ scattering with $k_t>p_T$, should be treated as the hard subprocess.

\subsection{The optimum scale to resum $(\alpha_s{\rm ln}(1/x){\rm ln}\mu^2)^n$ terms \label{sec:opt41}}
We use the formulae from appendix B of the \cite{CCH} paper
%MCFM and FONLL programmes \cite{MCFM,FONLL} 
to calculate the $gg\to Q\bar{Q}q$ matrix element in the high energy limit in order to find a scale,
\be
\mu_F~\equiv ~\mu_0~=~F*\sqrt{p^2_T+m^2_Q},
\ee
that nullifies the double-logarithmic NLO contribution: that is, to find a scale $\mu_0$ at which the DGLAP-induced contribution $(P_{\rm left}\otimes C^{(0)} +C^{(0)} \otimes P_{\rm right})$ 
%reproduces 
replaces the NLO correction calculated explicitly. 
Note, however, that calculating the NLO contribution of the diagram in Fig.~\ref{fig:toLO}, we have integrated over the momenta of other particles; in particular, over the transverse momentum, $-k_t$, of the light quark. Since we are going to consider the upper (heavy quark) box in Fig.~\ref{fig:toLO} as the `hard' subprocess, and would like to keep the DGLAP $k_t$ ordering, we put an additional cut  $-|k_t|<{\rm min}\{m_{TQ},m_{T\overline{Q}}\}$; otherwise the lower part of diagram (which may be either $qQ$ or $q\bar{Q}$  scattering) may have $k_t>m_T$, and would then be treated as the hard subprocess. Here $m_T=
\sqrt{m^2_Q+p^2_T}$.

\begin{figure} [h]
\begin{center}
\vspace{-0.0cm}
\includegraphics[clip=true,trim=0.0cm 0.0cm 0.0cm 0.0cm,width=9.0cm]{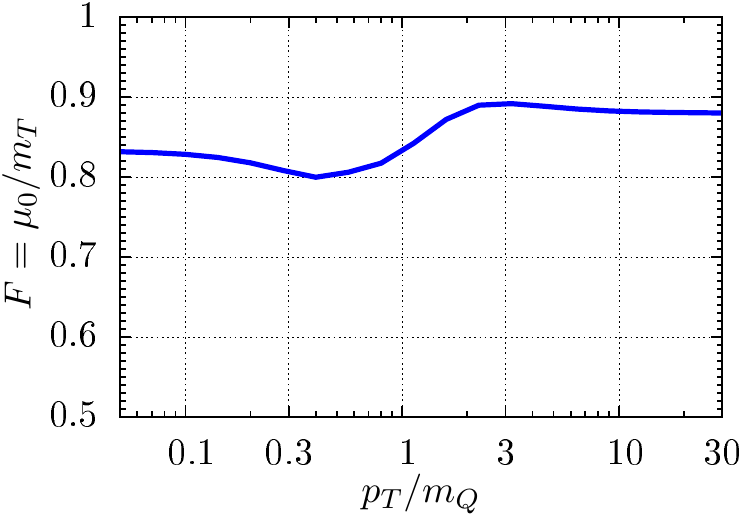}
\caption{\sf The optimal scale, $\mu_0=Fm_T$, as a function of $p_T/m_Q$. }
\label{fig:mu0}
\end{center}
\end{figure}

The values that we find for the `optimal' scale $\mu_0$
are presented in Fig.~\ref{fig:mu0} as the function of $p_T/m_Q$ ratio, where $p_T$ is the transverse momentum of the observed heavy quark. It turns out that the values of the optimal scale are close to the value $\mu^2_F=m^2_T\equiv p^2_T+m^2_Q$ that is used conventionally; that is $F=1$.   However, we now have a physics justification for the scale choice shown in Fig.~\ref{fig:mu0}, which to a good approximation is $\mu_0\simeq 0.85m_T$, that is $F\simeq 0.85$.

\begin{figure} [t]
%\vspace{-1.0cm}
\begin{center}
\includegraphics[width=0.5\textwidth]{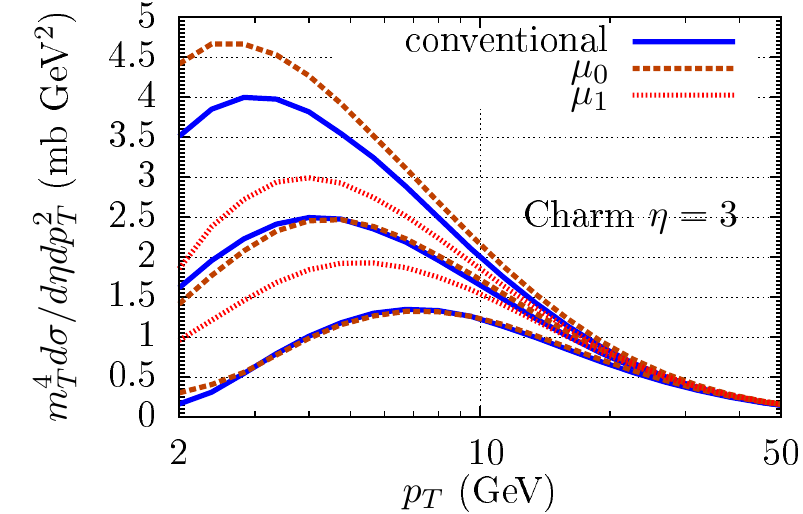}
\includegraphics[width=0.5\textwidth]{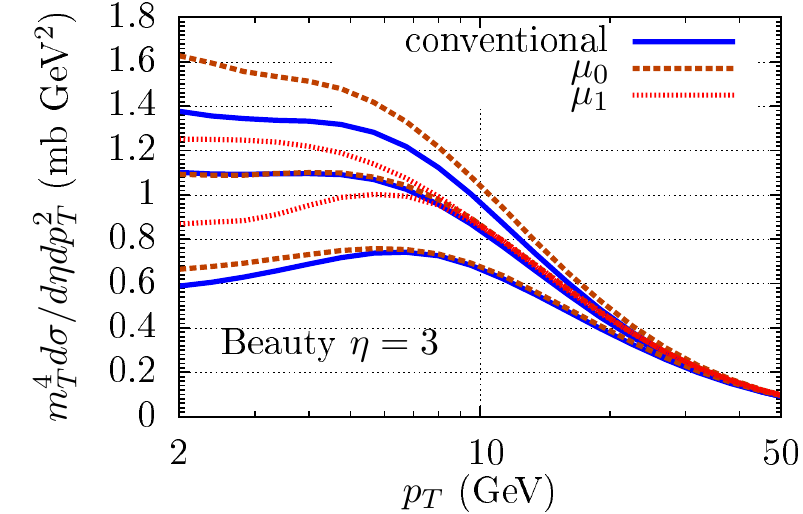}
\caption{\sf The scale dependence of the predictions of the cross section, $m_T^4{d\sigma}/{d\eta dp^2_T}$, for $c\bar{c}$ and $b\bar{b}$ production respectively, using the NLO CT14 parton set. The plot shows the scale variation $\mu_f =(2,1,0.5)m_T$ for three different procedures: (i) the conventional prediction (blue curves), (ii) resumming the ln$(1/x)$ contributions with $\mu_F=\mu_0\simeq 0.85m_T$ in (\ref{eq:stab1}) (dashed red curves), (iii) in addition resumming the $2\to 2$ contributions with $\mu_1=\mu_0$ in eqs.~(\ref{eq:stab11},~\ref{eq:22}) (dotted red curves).}   
\label{fig:xsec}
\end{center}
\end{figure}

 Now that we have the value of $\mu_0$, we can study the factorization scale,  $\mu_f$, dependence of the QCD predictions for $c\bar{c}$ and $b\bar{b}$ production. The results are shown in Fig.~\ref{fig:xsec} for a $Q$ quark of pseudorapidity $\eta=3$ -- typical of the LHCb experiment. 
The curves for the first two procedures\footnote{The third procedure is the subject of Section \ref{sec:2to2}.}, mentioned in the  caption of Fig.~\ref{fig:xsec}, are obtained from (\ref{eq:stab1}) setting $\mu_F$ (and $\mu_R$) equal to $\mu_0$, and then varying $\mu_f$ in the range $(m_T/2, 2m_T)$.
 We  use the CT14~\cite{Dulat:2015mca} PDFs as an example of a recent set of partons which have no negative gluon distributions and take the corresponding heavy quark masses:  $m_c=1.3$ GeV
and $m_b=4.75$ GeV. Subroutines from the MCFM~\cite{MCFM} and FONLL~\cite{FONLL} programmes were used for the computations.

For simplicity we take $F=0.85$, that is, $\mu_0=0.85m_T$, and make predictions for three different values of the factorization scale $\mu_f$, namely $\mu_f=(0.5,\ 1,\ 2)m_T$.
 The results are shown in Fig.~\ref{fig:xsec} by the dashed red curves.  We repeat the cross section prediction, but now use (\ref{eq:sta}) with the conventional choice $\mu_f=(0.5,\ 1,\ 2)m_T$, which gives the blue curves.
 Not surprisingly, since the optimum scale is close to the conventional choice $\mu_0= m_T$, the scale uncertainties are comparable.

Unfortunately we still have rather strong dependence of the predicted cross section on the choice of the value of $\mu_f$.   It is caused by the relatively low mass contribution coming mainly from the $2\to 2$ $(gg\to Q\bar{Q}$) component of $C^{(1)}$. This component does not contain a $\ln(1/x)$ dependence, but, at our  low scales, it is numerically large; it gives up to twice as  large a contribution as the LO one. Moreover, being convoluted with low-$x$ PDFs, which strongly depend on $\mu_f$, it produces a large scale uncertainty.

\begin{figure} [t]
%\vspace{-1.0cm}
\begin{center}
\includegraphics[width=0.6\textwidth]{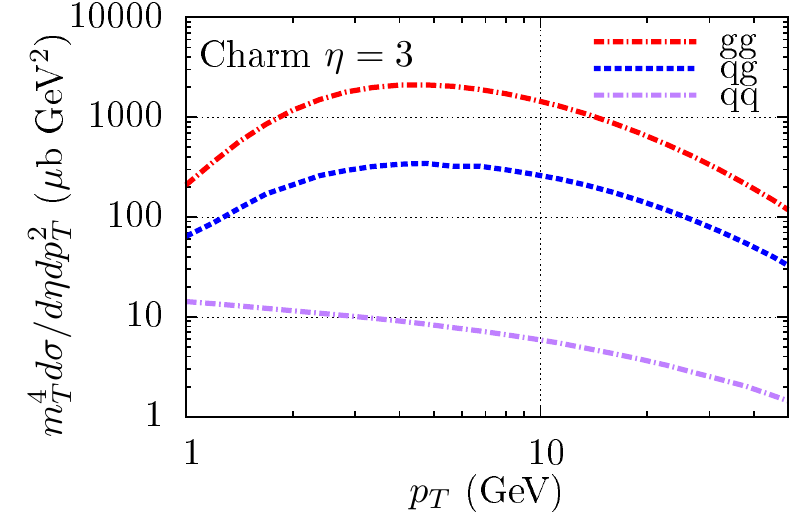}
\caption{\sf The $gg,~gq, q\bar{q}$  fusion contributions to $c\bar{c}$ production for a charm quark produced at pseudorapidity $\eta=3$ at in $pp$ collisions at 13 TeV}   
\label{fig:gg}
\end{center}
\end{figure}

\subsection{The optimum scale to resum the higher-order $2\to 2$ diagrams   \label{sec:2to2}}

In order to reduce the scale dependence it turns out to be important to fix the scale of the $2\to 2$ NLO contribution, that is to know the value of $\mu_1$ in the $2\to 2$ part of the second term on the right hand side of eq.(\ref{eq:stab11}). Strictly speaking to do this we have to know the NNLO expression. At the moment there exist only numerical NNLO result for $t$-quark pair production, see~\cite{top16} and references therein. Nevertheless we can extract some use from these calculations. As was demonstrated in \cite{forte} (see Fig. 6 for example) the corrections to the $2\to 2$ NLO contributions are mainly of 
Sudakov origin\footnote{Besides this there are, of course, the `renorm. group' corrections, which account for the possible variation of the value of $\mu_R$.}; -- these are `soft' corrections corresponding to a relatively small momentum transferred along an additional gluon. Such corrections does not change essentially the original kinematics of the $2\to 2$ subprocesses or the  dependence of the corresponding `hard' matrix element on the virtuality of incoming parton.   

Thus it looks reasonable to convolute these terms with the same PDFs as those used for the LO evaluation. This will provide the correct resummation of the higher-order DL terms, $(\alpha_s\ln \mu_F\ln(1/x))^n$ (with $n=2,3,...$) inside the incoming parton distributions. Referring to (\ref{eq:stab11}), it means that we may
argue that the $2\to 2$ part of the NLO coefficient function $C^{(1)}$ must be convoluted with partons taken at the scale $\mu_1=\mu_0$.   
In other words we write the cross section as
$$\sigma^{(0)}(\mu_f)+\sigma^{(1)}(\mu_f)~~=~~\alpha_s^2~[~{\rm PDF}(\mu_F)\otimes C^{(0)}\otimes {\rm PDF}(\mu_F)~~+$$
$$~~~~~~~~~~~+~~\alpha_s {\rm PDF}(\mu_F)\otimes C^{(1)}_{(2\to 2)}(\mu_F)\otimes {\rm PDF}(\mu_F)~~+$$
\be
\label{eq:22}
~~~~~~~~~~+~~\alpha_s {\rm PDF}(\mu_f)\otimes C^{(1)}_{(2\to 3)}(\mu_F)\otimes {\rm PDF}(\mu_f)~]\ ,
\ee
with $\mu_F=\mu_0$ and $\alpha_s(\mu_R)=\alpha_s(\mu_0)$, where we have divided the $C^{(1)}$ correction into two terms $C^{(1)}=C^{(1)}_{(2\to 2)}+C^{(1)}_{(2\to 3)}$, with only the second term evaluated at the residual factorization scale $\mu_f$.
The corresponding results, calculated from (\ref{eq:22}), are shown in Fig.~\ref{fig:xsec} by the dotted red curves.
 We see that the remaining $\mu_f$ dependence is much reduced. We consider this observation as a strong argument in favour of the possibility of using open charm or beauty data to constrain the low $x$ partons at the scale $\mu_f=0.85m_T$.
 
 Since the major contribution to $c\bar{c}$ and $b\bar{b}$ production comes from gluon-gluon fusion (see Fig.~\ref{fig:gg}), including  these data in global parton analyses will allow a better study of the gluon low-$x$ behaviour, and hence to strongly diminish the present uncertainty observed in this region.

\section{Azimuthal cut to reduce optimal scale for $b\bar{b}$ events  \label{sec:azim} }
In the case of  open $b\bar{b}$  production the optimal scale is rather large; typically  $\mu^2_0>30$ GeV$^2$. On the other hand, the main uncertainties in the gluon PDF are observed at much lower scales $\sim $ 2 - 4 GeV$^2$. One possibility to reduce the scale at which the process probes the partons is to observe both heavy quarks (i.e. both the quark and the antiquark), and then to select the events where the transverse momentum of the pair is small. This proposal was discussed in~\cite{kt} (and in~\cite{DY} for Drell-Yan pair production). Unfortunately, the {\it transverse momenta} of $B$ mesons can only be measured for a few particular decay modes, and the product of the branching ratios for the two $B$ mesons is small. It means that we do not have sufficient statistics.

Another idea was proposed by Alexey Dzyuba\footnote{We thank Alexey Dzyuba of the Petersburg Nuclear Physics Institute for this idea.}. As a rule the vertex of $B$ meson decay can be observed experimentally, and it is possible to measure the {\it azimuthal angle}, $\phi$, between the two heavy mesons. That is, we  may select $B\bar{B}$ events with good coplanarity.
In such a case  the transverse momenta of the incoming partons must be small, otherwise the coplanarity will be destroyed. In other words, for events with a small $\Delta\phi=\pi-\phi$ we deal with lower scale partons. For example, in Table~\ref{tab:A} we show
the optimal scale $\mu_0(\Delta\phi)$ calculated for events with $\Delta\phi<\phi_0$ corresponding to the LHCb rapidity interval $2<y<4.5$.
As expected,  for low $\phi_0$ we have $\mu_0\propto\phi_0$. For instance, for $b\bar{b}$ production with
 $\Delta\phi<10^o$ one can probe gluons at a rather low scale, namely  $\mu\simeq 1.5$ GeV.
\begin{table}
\begin{center}
\begin{tabular}{|c|c|c|c|c|c|c|c|}\hline
$\phi_0$ & 0.0875 &  0.175 &  0.263  & 0.350 &  0.438 & 0.525 & radians\\
          &     $5^o$     &   10 &     15   &  20  &    25   &  30  & degrees\\ \hline
  $\mu_0/m_b$ &  0.17  &  0.33 & 0.46 &  0.58 &  0.69 &  0.78 &  \\ 
%\hline
 $\mu_0$ & 0.83 & 1.57 & 2.20 & 2.76 & 3.26 & 3.70 & GeV\\ \hline
\end{tabular}\\
\caption{\sf The optimal factorization scale, $\mu_0$, corresponding to a coplanarity cut $\Delta\phi<\phi_0$ for events with both heavy quarks in the rapidity interval $2<y<4.5$. The cross section is integrated over the heavy quark transverse momenta $p_T$.} 
\label{tab:A}
\end{center}
\end{table}

\begin{figure} [h]
%\vspace{-1.0cm}
\begin{center}
\includegraphics[width=0.5\textwidth]{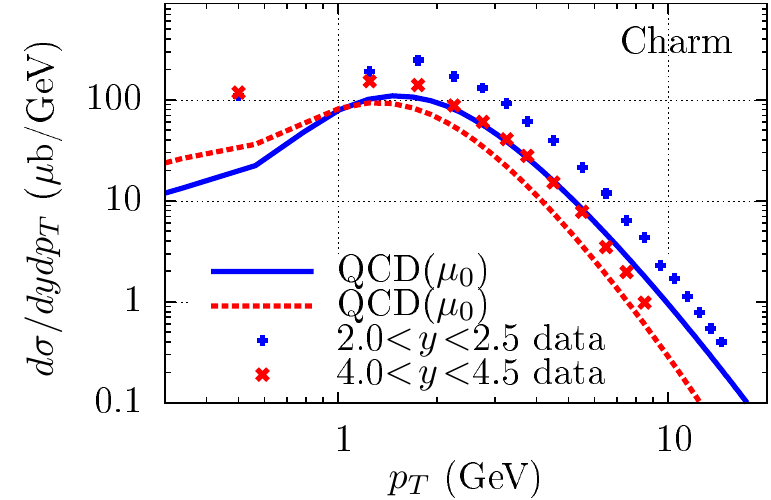}
\includegraphics[width=0.5\textwidth]{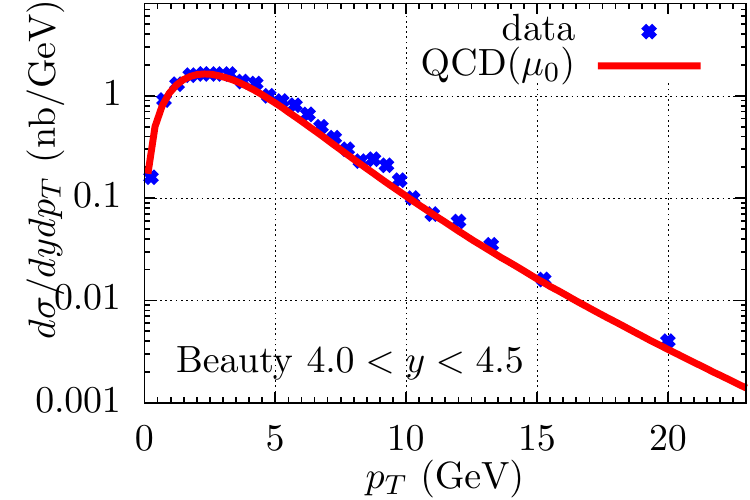}
\caption{\sf The QCD predictions for the cross section for heavy meson $(D^+,B^+)$  production  compared with LHCb data \cite{LHCbcc13,LHCbbb}, as a function of the $p_T$ of the heavy meson.   In the upper plot the blue (red) curves and data points, taken at $\sqrt{s}$=13 TeV, correspond to the $D^+$ rapidity bins $2<y<2.5~ (4<y<4.5)$ respectively; whereas the lower plot corresponds to $B^+$ rapidity in the interval $4<y<4.5$ for a collider energy of $\sqrt{s}$=7 TeV. CT14 NLO PDFs \cite{Dulat:2015mca}  are used.  The optimal factorization scale is taken $\mu_F=\mu_0=\mu_1=0.85m_T$; and the renormalization scale is taken to be $\mu_R=\mu_F$, see Section~\ref{sec:renorm}.}   
\label{fig:data}
\end{center}
\end{figure}

\section{Comparison with $c\bar{c}$ and $b\bar{b}$  data   \label{sec:data} }

Now that we have the optimal factorization scale, $\mu_0\simeq 0.85 m_T$, we can make an exploratory comparison with the existing LHC data for open single inclusive heavy-flavour production.
To compare with the data we use the subroutines from MCFM and FONLL programmes~\cite{MCFM,FONLL}.  
The QCD description of the present data in the low-$x$, low-$\mu$ domain is shown in Fig.~\ref{fig:data}. As an example, we consider just the $D^+$ ($B^+$) meson cross sections
using the probabilities of the quark to meson transition $P(c\to D^+)=0.25$ (see, for example, \cite{PDG2008} p.208) and $P(b\to B^+)=0.4$ (see \cite{LHCbbB} and \cite{PDG} p.63).
%{\bf Emmanuel, what numbers were used here. Please, include}\\
 We account for the fact that the $D/B$ meson momentum is less than that of the parent quark by making the assumption that $p_D\sim 0.75p_c$ and $p_B\sim 0.9p_b$ (see~\cite{PDG,Cacc2}). 
That is, we calculate the meson cross sections as
\be
\frac{d\sigma_{D^+}}{dydp_{T,D}}~=~0.25\frac{d\sigma_c(p_{T,c}=p_{T,D}/0.75)}
{dydp_{T,c}}\frac 1{0.75} \ ,
\ee
\be
\frac{d\sigma_{B^+}}{dydp_{T,B}}~=~0.4\frac{d\sigma_b(p_{T,b}=p_{T,B}/0.9)}{dydp_{T,b}}\frac 1{0.9}\ ,
\ee
where the last factor (1/0.75 or 1/0.9) accounts for the ratio of the $dp_{T,D(B)}$ and $dp_{T,c(b)}$ intervals.

It is seen that the  QCD predictions obtained using the 'optimal' factorization scale and the central values of CT14 NLO partons underestimate the LHCb  $c\bar{c}$ data. Note, however, there are large uncertainties in the behaviour of the low-$x$ gluon distributions obtained from the global parton analyses. This uncertainty may be reduced for the NLO partons~\footnote{Formally at the NLO level we do not account for the NNLO corrections.} by including the open charm/beauty data in the global analysis and using the `optimal' scale to calculate the corresponding cross sections.
Of course, there are also the uncertainties due to higher $\alpha_s$ order contributions not included into the calculations.
 When the NNLO formulae become available it will be possible to extend our procedure and to include open charm data into the NNLO global parton analyses.

\section{Conclusion   \label{sec:conc} }
We have calculated the `optimal' factorization scale, $\mu_0$, which allows a resummation of the higher-order $\alpha_s$ corrections, enhanced at high energies by the large $\ln(1/x)$ factor; that is to resum the double logarithmic, $(\alpha_s\ln\mu_F^2\ln(1/x))^n$, terms and move them into the incoming parton distributions.
The result is given in Fig.~\ref{fig:mu0}.  It is essentially
\be
\mu_F~=~\mu_0~\simeq ~0.85\sqrt{p^2_T+m_Q^2}
\label{eq:A8}
\ee
for single open inclusive heavy quark production, where $p_T$ is the transverse momentum of the observed heavy quark.

We also considered the case when the azimuthal angle, $\phi$, between the heavy quark and the antiquark can be measured. We showed that by selecting events with small $\Delta\phi=\pi-\phi$ we are able to probe smaller factorization scales $\mu_0$.  This is an advantage for $b\bar{b}$ production: compare the results of Table~\ref{tab:A} with eq,~(\ref{eq:A8}). The disadvantage is that the rate is smaller for such events, even though we do not require that the transverse momentum of both the heavy quarks are measured. 

The choice $\mu_F=\mu_0$ reduces the uncertainty of the perturbative QCD calculations. It will allow LHC data on $c\bar{c}$ and $b\bar{b}$ production to be included in global parton analyses to constrain the behaviour of the gluon distribution
 in the region of very small $x$ and low scale, equal to $\mu_0$,  where the uncertainties of the
%extrapolation based on the 
present global parton analyses are especially large.

\section*{Acknowledgments}

We thank Keith Ellis for valuable discussions. MGR and EGdO thank the IPPP at the University of Durham for hospitality. This work 
was supported by the RSCF grant 14-22-00281 for MGR and by Capes and CNPq (Brazil) for EGdO.

%\bibitem{kt} 
%The small x gluon and b\bar{b} production at the LHC 
 % E.G. de Oliveira, A.D. Martin, M.G. Ryskin,  Eur.Phys.J. {\bf C71} (2011) 1727 

%\bibitem{DY}  
%  Improving the Drell-Yan probe of small x partons at the LHC via a k_t cut 
%  E.G. de Oliveira, A.D. Martin, M.G. Ryskin, Eur.Phys.J. {\bf C73} (2013) 
%no.3, 2361 
\end{document}